\documentclass[%
 reprint,
superscriptaddress,
 amsmath,amssymb,
 aps,
prl
]{revtex4-2}

\usepackage{graphicx}
\usepackage{dcolumn}
\usepackage{bm}

\usepackage{lipsum}
\usepackage{color}

\begin{document}

\preprint{APS/123-QED}

\title{Freezing a rivulet}

\author{Antoine Monier}
\affiliation{Sorbonne Universit\'e, CNRS, UMR 7190, Institut Jean Le Rond $\partial$'Alembert,  F-75005 \ Paris, France}%

\author{Axel Huerre}%
 \email{axel.huerre@gmail.com}
 \affiliation{Laboratoire d'Hydrodynamique (LadHyX), UMR 7646 CNRS-Ecole Polytechnique, IP Paris, 91128 Palaiseau CEDEX, France}%
 
 \author{Christophe Josserand}
\affiliation{Laboratoire d'Hydrodynamique (LadHyX), UMR 7646 CNRS-Ecole Polytechnique, IP Paris, 91128 Palaiseau CEDEX, France}%

\author{Thomas S\'eon}
\affiliation{Sorbonne Universit\'e, CNRS, UMR 7190, Institut Jean Le Rond $\partial$'Alembert,  F-75005 \ Paris, France}%

\date{\today}

\begin{abstract}
We investigate experimentally the formation of the particular ice structure obtained when a capillary trickle of water flows on a cold substrate. We show that after a few minutes the water ends up flowing on a tiny ice wall whose shape is permanent. 
We characterize and understand quantitatively the formation dynamics and the final thickness of this ice structure. 
In particular, we identify two growth regimes. First, a 1D solidification diffusive regime, where ice is building independently of the flowing water.
And second, once the ice is thick enough, the heat flux in the water comes into play, breaking the 1D symmetry of the problem, and the ice ends up thickening linearly downward.
This linear pattern is explained by considering the confinement of the thermal boundary layer in the water by the free surface.
\end{abstract}

\maketitle

The partial freezing of Niagara falls and the cancellation of thousands of flights during the cold snap of winter 2019 are a few examples of the disturbances caused by extreme weather events. 
Indeed, the accretion of ice on super-structures such as planes \cite{Cebeci2003,Cao2018, Baumert2018}, power-lines \cite{Laforte1998}, bridge cables \cite{Liu2019} or wind turbines \cite{Wang2017} can have dramatic consequences. 
Nowadays, the main strategy to prevent most of these undesirable effects is to develop anti-icing surfaces \cite{Antonini2011,Kreder2016}, but new paradigms could emerge from a better understanding of the freezing dynamics in complex configurations.
When water flows on a cold surface for example, the resulting ice structure is reminiscent from the manifold interaction between the heat transport and the flow \cite{Davis2006}.
The presence of a free-surface is also determinant in these problems, resulting in the apparition of a tip on frozen sessile drops \cite{Marin2014}, or in the explosion of droplets cooled from the outside-in \cite{Wildeman2017} in static conditions. 
Freezing of capillary flows can consequently reveal a very rich behavior \cite{Moore2017, Herbaut2019} as in the formation of icicles \cite{Chen2011} or ice structures following drop impact on cold surfaces \cite{Thievenaz2019}.

\begin{figure*}
\centering
\includegraphics[width=1\textwidth]{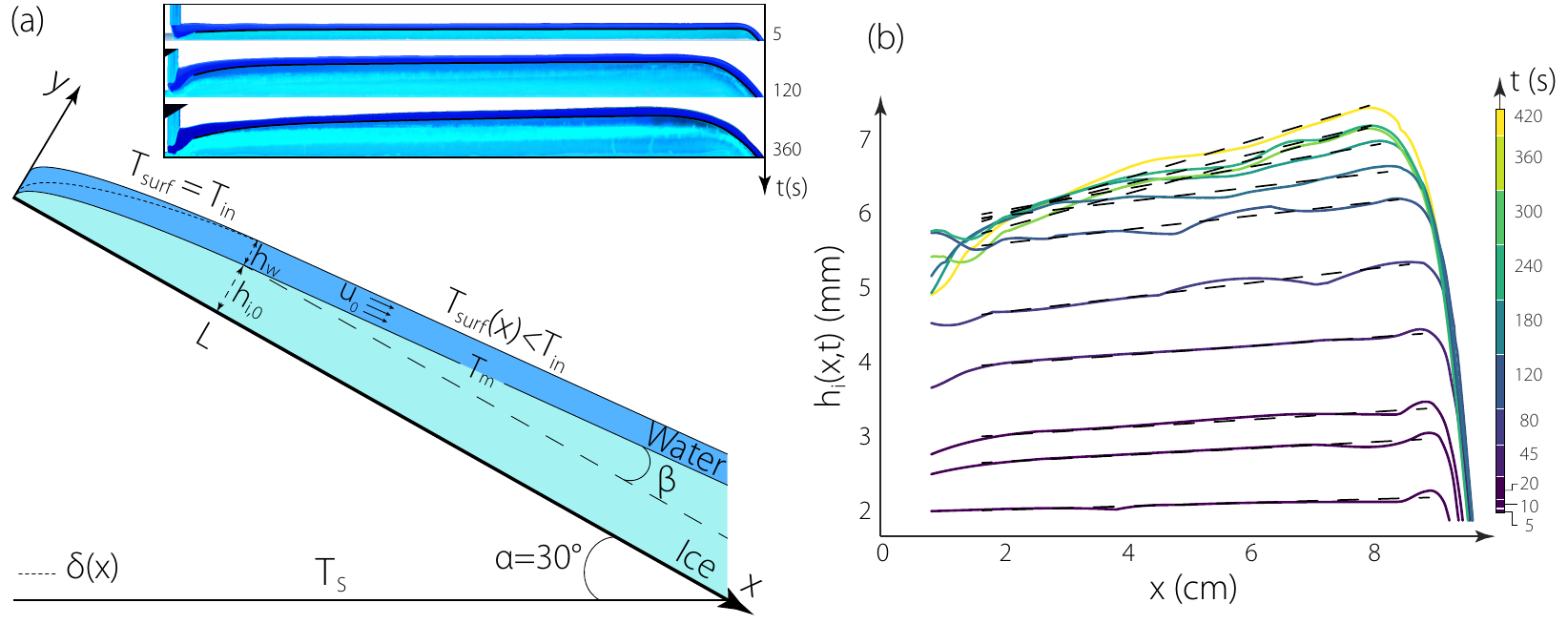}
\caption{(a) Schematic view of the freezing rivulet. The dotted line, $\delta(x)$, illustrates the thermal boundary layer. The inset presents an experimental sequence of three images.  The time origin is taken when the water starts flowing over the cold substrate. A black line is added between the water and the ice to highlight the interface.
(b) Profiles of the ice layer at different times for $T_{\rm in}$=10$^{\circ}$C and $T_{\rm s}$=-36$^{\circ}$C. This experiment is the same as shown on the inset of (a). The permanent regime can be described by $h_{\rm max}(x)=h_{\rm i,0}+\beta x$.}
\label{Fig1}
\end{figure*}

In this Letter, we investigate experimentally the freezing of a capillary water river, the so-called rivulet \cite{Towell1966, LeGrand-Piteira2006, Daerr2011} (see Fig.~\ref{Fig1}), flowing over a cold substrate.
We show for the first time that the growing ice structure reaches a static shape after few minutes. The water then flows on a tiny ice wall that thickens downward, an observation we quantitatively explain considering the confinement of a thermal boundary layer.


The experiment consists in flowing distilled water dyed with fluorescein at 0.5$\,$g.L$^{-1}$ along a cold aluminum block of 10$\,$cm long, with an inclination of $\alpha=30^{\circ}$ to the horizontal. The temperature of the injected water $T_{\rm in}$ ranges from 8 to 35$^{\circ}$C, see Fig.~\ref{Fig1}(a). The water is injected through a needle (inner diameter 1.6$\,$mm) at a flow-rate $Q=20\,$mL.min$^{-1}$, such that there is no meander at room temperature \cite{LeGrand-Piteira2006}. A straight water rivulet is then formed \cite{Towell1966}, with a typical width of $2\,$mm, a thickness of $h_{\rm w}=800\,\mu$m, and a characteristic velocity of the buoyant flow $u_0\approx 10\,$cm.s$^{-1}$. As the Reynolds number of the flow is sufficiently small ($Re=u_0h_{\rm w}/\nu=80$), the flow is laminar and mass conservation implies that the liquid layer thickness $h_{\rm w}$ is constant \cite{Towell1966}.
The temperature of the aluminum substrate $T_\text{s}$ is set by plunging the block in liquid nitrogen for a given amount of time so that it ranges from $-9$ to $-44^{\circ}$C. Experiments realized with substrate temperatures below $-44^{\circ}$C consistently lead to the fracture  \cite{Ghabache2016} or the self-peeling \cite{Ruiter2018} of the ice and are not considered here.
Upon contact with the cold substrate, the water freezes and an ice layer grows while the water continues to flow on top, as shown on the sequence of snapshots of inset in Fig.~\ref{Fig1}(a) and in the Sup. Mat. movie.
During that process, the fluorescein concentrates between the ice dendrites, causing self-quenching and fluorescence dimming in the ice \cite{Marcellini2016}. This allows us to clearly distinguish between the ice and the water layers under UV light.   
The ice layer thickness $h_{\rm i}(x,t)$ is then measured using a Nikon D800 camera recording from the side at 30 fps.


Figure~\ref{Fig1}(b) presents the ice layer profile along the direction of the flow ($x=0$ at the needle) at different times for $T_{\rm in}$=10$^{\circ}$C and $T_{\rm s}$=-36$^{\circ}$C. 
The analysis is restricted to the middle of the plate (x $\in$ [1,8]$\,$cm) to avoid input and output influences.
 At early times, the ice layer grows homogeneously along the plane and the successive profiles are parallel to the substrate. After that, the ice layer continues to grow but not uniformly: its thickness increases along the plane. Finally, the ice layer stops growing and the system reaches a permanent regime consisting of a static ice structure, of thickness $h_\text{max}$, on top of which a water layer is flowing. The final shape of the ice can be well described by a line of slope $\beta$ as illustrated by the dashed lines in Fig.~\ref{Fig1}(b):  $h_{\rm max}(x)=h_{\rm i,0}+\beta x$, with $\beta$ varying in our experiments between $0$ and $4^\circ$.
The presence of such a permanent regime can be understood  qualitatively by considering the thermal fluxes at the ice-water interface. Because the ice acts as a thermal diffusive layer between the plate and the liquid, the cooling flux - through the ice layer - diminishes as the ice layer thickens and the temperature gradient decreases. On the other hand, since the flowing water is dispensed at constant temperature on the forming ice layer, the heat flux brought to the system is constant. Consequently, an equilibrium is reached when the ice layer thickness is such that both fluxes balance. The aim of the present paper is therefore to characterize and understand quantitatively the formation dynamics and the final thickness of these ice structures.


\begin{figure}
\centering
\includegraphics[width=0.5\textwidth]{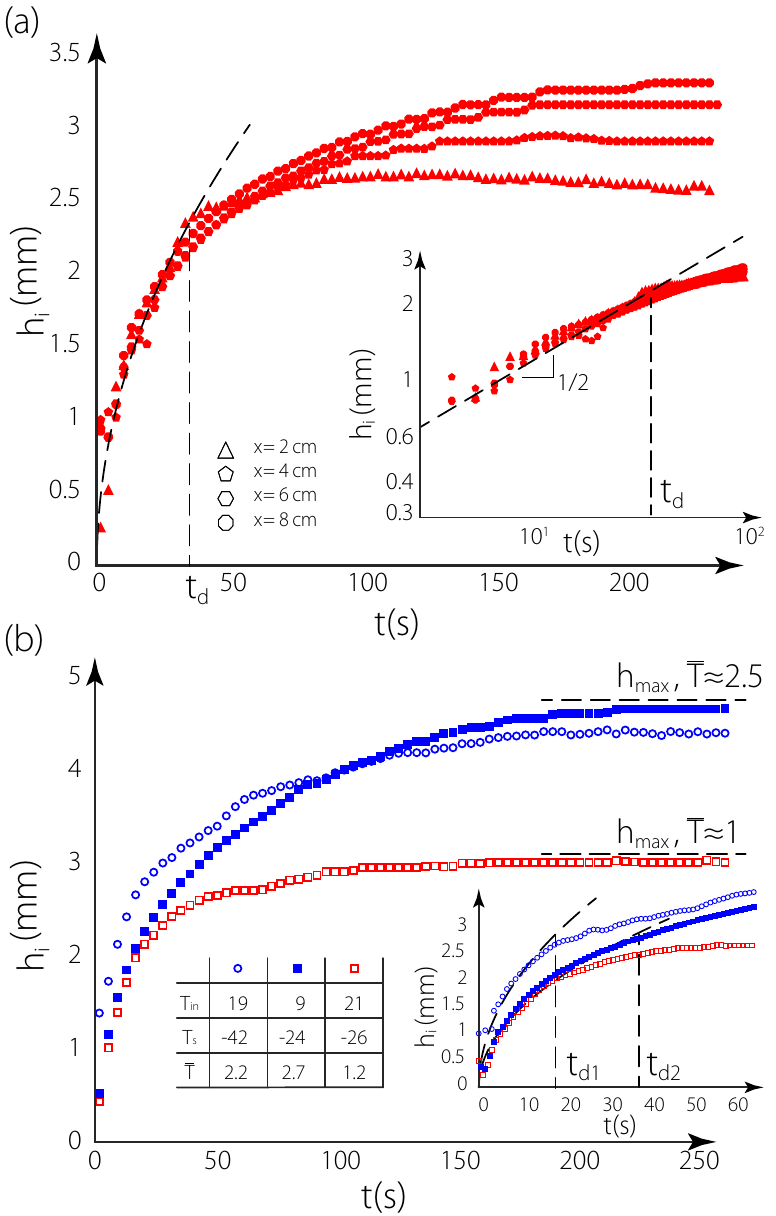}
\caption{Time evolution of the ice layer thickness. (a) Ice layer thickness $h_{\rm i}(t)$ at different positions along the plane : x=2, 4, 6, 8$\,$cm. After a time $t_{\rm d}$, the ice growth enters a new regime where it depends on the position on the plane. Inset: loglog representation of $h_{\rm i}(t)$. For t$<t_{\rm d}$, all the curves collapse on the dashed line of slope 1/2. (b) Ice layer thickness at x=6$\,$cm for different thermal parameters.
Inset: zoom at early times with indication of the different $t_{\rm d}$.}
\label{Fig2}
\end{figure}

\begin{figure}
\centering
   \includegraphics[width=0.491\textwidth]{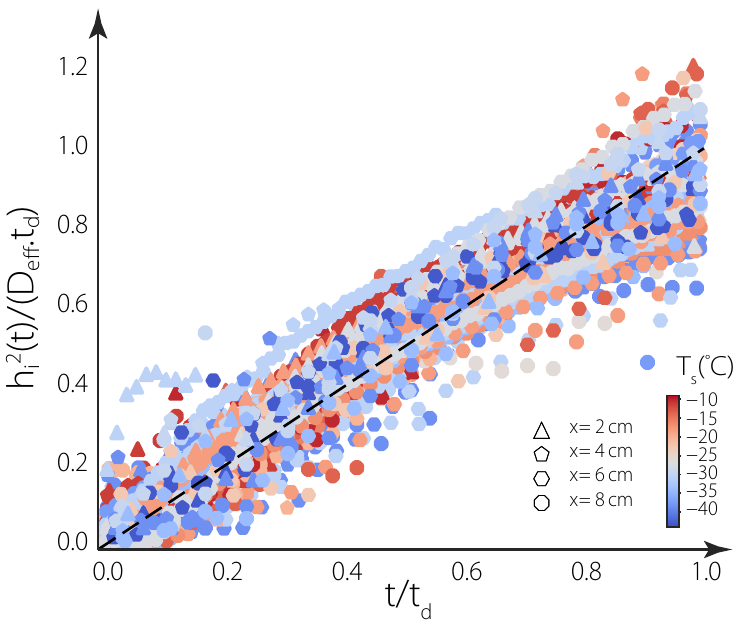}
\caption{Rescaled ice thickness $h_{\rm i}^2(t)/(D_{\rm eff} t)$, $t\in [0,t_\text{d}]$ as a function of the non-dimensional time $t/t_{\rm d}$. $D_{\rm eff}$ is computed from \cite{Thievenaz2019} and $t_{\rm d}$, the time where the spatially homogeneous growth ends, is experimentally determined.}
\label{Fig3}
\end{figure}

Figures~\ref{Fig2} and \ref{Fig3} describe the temporal evolution of the ice thickness at different positions along the plane.
Figure~\ref{Fig2}(a) shows the increase of $h_{\rm i}(x,t)$ in time for four different positions along the x-axis ($x=2$, $4$, $6$, $8\,$cm).
In each case, the ice layer grows rapidly at short time and then converges after a few minutes to its maximum $h_{\rm max}(x)$.
Noticeably, all the curves superimpose at early times indicating that the dynamics there does not depend on the position along the plane.
Moreover, in this initial regime, the curves follow reasonably the power-law of exponent 1/2 represented by the dashed line, which is confirmed by the log-log plot of inset of Fig.~\ref{Fig2}(a). 
This suggests an initial common diffusive growth dynamics and defines the characteristic time $t_{\rm d}$ of this early regime, precisely when the profiles depart from the square-root behavior.

Figure~\ref{Fig2}(b) presents the dynamics of the ice layer at $x=6\,$cm for three different values of the temperature couple: $T_{\rm in}$ (injected water) and $T_{\rm s}$ (substrate), indicated on the legend of the graph.
The ice layer formation depends on the water and substrate temperatures but, remarkably, different values of both $T_{\rm in}$ and $T_{\rm s}$ lead to the same $h_{\rm max}$ (blue markers). 
The good variable to consider is then the reduced temperature $\bar{T}=(T_{\rm m}-T_{\rm s})/(T_{\rm in}-T_{\rm m})$ where $T_{\rm m}=0^{\circ}$C is the melting temperature. Indeed, the blue curves have a similar $\bar{T}$, different from the red curve's one. 
The static ice layer thickness ($h_\text{max}$) increases with $\bar{T}$, when the substrate and/or the liquid get colder, as expected. 
The inset of Fig.~\ref{Fig2}(b) is a zoom at early times, emphasizing the role of the temperature in the initial growth.
Keeping the substrate temperature $T_{\rm s}$ constant (square symbols), the dynamics is not modified by changing the water temperature $T_{\rm in}$. However, the regime holds longer when the water is colder (blue squares) : $t_{\rm d2} > t_{\rm d1}$.
Conversely, for different $T_{\rm s}$ and same $T_{\rm in}$ (empty symbols), the dynamics is different but $t_{\rm d1}$ is the same. The dynamics seems thus to be controlled by the substrate temperature with no influence of the water temperature; the latter controlling the length of this initial regime.

To understand this short time behavior of the ice layer growth, we model it by using 
the Stefan condition, where the latent heat produced by the ice formation results from the difference between the heat fluxes through the ice and the water. It reads here:
\begin{equation}
\rho_{\rm i} \mathcal{L} \partial_{t} h_{\rm i}=\lambda_{\rm i} \partial_y T \left(x,h_{\rm i}^-,t \right)-\lambda_{\rm w} \partial_{y} T\left(x, h_{\rm i}^+, t\right),
\label{eq:Stefan}
\end{equation}
where $\rho_{\rm i}$ is the ice density, $\lambda_{\rm i,w}$ are the heat conductivities of the ice and water respectively, and $\mathcal{L}$ the latent heat of solidification.
Initially, when $h_{\rm i} \ll h_{\rm w}$, the flux through the ice, which scales as $\lambda_{\rm i} (T_{\rm m}-T_{\rm s})/h_{\rm i}$, is dominant compared to the flux through the water.
This explains why the dynamics is independent of $x$ in this first regime as the temperature is constant along the surface of the substrate. 
This dynamics at short time corresponds to the classical one dimensional problem of the growth of an ice layer when a liquid is suddenly put in contact with a substrate at a uniform subfreezing temperature, usually known as the \textit{classical Stefan problem} \cite{Stefan1891, Rubinstein1971}.
In this situation, the front follows a diffusive dynamics: $h_{\rm i} (t)=\sqrt{D_{\rm eff} \,t}$, where the coefficient $D_{\rm eff}$ is solution of a transcendental equation that involves $T_{\rm s}$, $\mathcal{L}$ and the ice and aluminum thermal coefficients.

In Fig.~\ref{Fig3}, we plot the rescaled ice thickness $h_{\rm i}^2(t)/(D_{\rm eff} t_{\rm d})$, $t\in [0,t_{\rm d}]$, where $D_{\rm eff}$ is calculated using a refined model of the classical Stefan problem \cite{Thievenaz2019}, versus the non-dimensional time $t/t_{\rm d}$. 
The graph presents the results of 31 experiments and 4 different positions along the plane, where $T_{\rm s}$ varies from -9 to -44$^{\circ}$C and $T_{\rm in}$ from 8 to 35$^{\circ}$C.
All the data collapse on a line of slope 1, confirming that at short time the ice growth is only determined by the heat transfer toward the substrate without influence of the water flow.

This regime ends when the heat flux in the water is no more negligible and becomes comparable to the heat flux in the ice. $t_{\rm d}$ can therefore be estimated by balancing the two heat fluxes involved in the Stefan equation (\ref{eq:Stefan}), leading to the scaling $\lambda_{\rm i} (T_{\rm m}-T_{\rm s})/\sqrt{D_{\rm eff}t_{\rm d} }\sim \lambda_{\rm w} (T_{\rm in}-T_{\rm m})/h_{\rm w}$. 
Moreover, for the present experiments, the heat flux in the ice is smaller than the latent heat, so the Stefan number ($St=c_{\rm p,i} (T_\text{m}-T_\text{s})/\mathcal{L}$, where $c_{\rm p,i}$ is the heat capacity of the ice) is smaller than 1 and $D_{\rm eff}$ is proportional to $St^2$ \cite{Thievenaz2019}.
Remarkably, the terms $(T_{\rm m}-T_{\rm s})$ cancel out, showing that in this regime $t_\text{d}$ depends on the water temperature $T_\text{in}$  but is independent of the substrate temperature $T_{\rm s}$, as observed in our experiments (see inset of Fig.~\ref{Fig2}(b)).\\

For times larger than $t_{\rm d}$, the heat exchange between the ice and the water is now influenced by the water flow that advects heat along the slope. Therefore, the dynamics depends on the $x$ axis and becomes much more complex but, as shown before, the ice structure eventually adopts a static shape. In the following the paper focuses on this steady regime.
As sketched in Fig.~\ref{Fig1}(a), the steady ice structure can be spatially divided into two zones for all the experiments.
In the first one, close to the water inlet, the ice shape is expected to adopt a square-root function of $x$ \cite{Hirata1979}. 
This behavior is valid until $x=L$, leading to an ice thickness $h_\text{max}(L) = h_{\rm i,0}$. The second zone, for $x>L$ corresponds to the linear shape illustrated in Fig.~\ref{Fig1}(a).


To describe this steady regime, both the heat equation in the ice and in the water layers are considered.
In the ice, the temperature field is stationary, it follows $\partial_{yy} T=0 $ and is thus simply linear with $y$: $T(x,y)=T_{\rm s}+(T_{\rm m}-T_{\rm s})y/h_{\rm max}(x)$.
In the water, the quasi-parallel approximation ({\it i.e.} $|\partial_x| \ll |\partial_y|$) is considered and we use a representative flow field with a constant velocity $u_0$ (plug flow) such that the advection-diffusion equation to solve reads:
\begin{equation}
    u_0 \,\partial_x T=D_{\rm w}\partial_{yy} T,
    \label{HeatEquationWater}
\end{equation}
where $D_{\rm w}$ is the heat diffusion coefficient in the water. 
These equations are coupled through the boundary conditions that impose the continuity of the temperature ($T=T_{\rm m}$) and of the heat fluxes at the ice/water interface, $y=h_{\rm max}(x)$. Indeed, since no more phase change occurs the Stefan condition (Eq. (\ref{eq:Stefan})) leads to the equality of the fluxes at the interface.
Moreover, no heat flux towards the surrounding air is considered, leading to $\partial_y T(x,h_{\rm max}+h_{\rm w})=0$ and the water temperature is imposed at the inlet: $T(0,y)=T_{\rm in}$.
But as this latter condition is incompatible with the interface imposed temperature $T_{\rm m}$, a thermal boundary layer has to establish from the inlet \cite{Prandtl1904}. 
The thickness of this boundary layer $\delta(x)$ (see Fig.~\ref{Fig1}(a)) is given by $\delta(x)\simeq 4.2\,\sqrt{D_{\rm w} x/u_0}$ for a plug flow\cite{Schlichting2016} .

Interestingly, contrary to former studies \cite{Lapadula1966, Beaubouef1967, Hirata1979}, this thermal boundary layer cannot develop infinitely since it is bounded by the rivulet thickness $h_{\rm w}$. It only exists for small $x$ such that $\delta(x) <h_{\rm w}$, leading to $x\lesssim 0.06 \,h_{\rm w}^{2} \,u_{0}/D_{\rm w}\simeq 2 \,$cm, constant for all the experiments. 
This theoretical value is compatible with the one experimentally measured, $L$, defined as the length where the linear zone begins (see Fig.~1 in Sup. Mat). 

\begin{figure}
\centering
\includegraphics[width=0.5\textwidth]{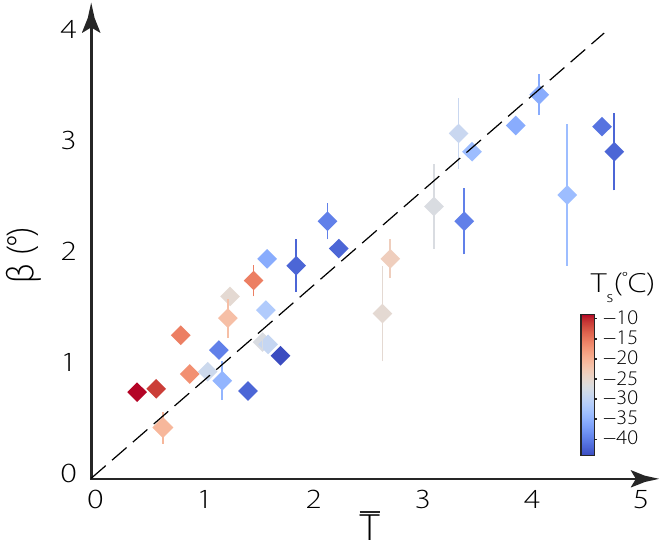}
\caption{Slope of the ice layer $\beta$ as a function of the rescaled temperature $\bar{T}=\frac{T_{\rm m}-T_{\rm s}}{T_{\rm in}-T_{\rm m}}$. The dashed line is a linear fit of the data $\beta = 0.77 \ \bar{T}$.}
\label{Fig4}
\end{figure}

The linear ice structure starts for $x>L$ when the thermal boundary layer becomes constrained by the free surface. Here, the water surface temperature $T_{\rm surf }(x)$, at $y=h_{\rm max}(x)+h_{\rm w}$ is not constant anymore, it decreases along $x$.
The temperature field in the water can be written using a classical self-similar ansatz:
\begin{equation}
 T(x,y)=T_{\rm m}+\left(T_{\rm surf }(x)-T_{\rm m}\right)F\left(\frac{y^{*}}{h_{\rm w}}\right),   
\end{equation}
where $F$ is a function solution of the full problem (Graetz solution \cite{Graetz1885}) and $y^*=y-h_{\rm max}(x)$ is the relative position in the water layer.
Plugging this ansatz into the heat equation (Eq. (\ref{HeatEquationWater})) gives:
\begin{equation}
 u_{0} \frac{d T_{\rm surf }(x)}{d x} = -a\, D_{\rm w} \frac{T_{\rm surf }(x)-T_{\rm m}}{h_{\rm w}^{2}} \ ,   
\end{equation}
 where we made the slender body approximation $h_{\rm max}'(x) \ll 1$ and $a$ is a constant.
The solution of this equation involves an exponential function of $x$ that can be linearized so that 
$T_{\rm surf}\left(x^*\right)\sim T_{\rm in}-a\,\left(T_{\rm in}-T_{\rm m}\right) x^{*}$, with $x^* = (x-L)/L$ for $a\,x^{*}\ll 1$.
The ice-water boundary condition (Eq. (\ref{eq:Stefan}) with $\partial_t h_\text{max} =0$) leads then to the expected linear ice profile in this domain:
\begin{equation}
    h_{\rm max}(x^*)\sim h_{\rm w}\frac{\lambda_{\rm i}}{\lambda_{\rm w}}\frac{T_{\rm m}-T_{\rm s}}{T_{\rm in} - T_{\rm m}}(1+a\,x^*) \ .
\label{eq:slope}
\end{equation}

This calculation predicts, in particular, that the slope of the ice structure $\beta$ should vary linearly with the reduced temperature $\bar{T}$. 
Figure~\ref{Fig4} shows precisely the experimental values of $\beta$, measured for all of our experiments with different water and substrate temperatures, as a function of $\bar{T}$.
The color-code corresponds to experiments performed with different $T_\text{s}$. The dashed line is a linear fit of the data, showing a good agreement with the prediction of Eq. (\ref{eq:slope}). The fit gives $a=0.04$, justifying our analysis as long as  $x<25\,$cm. Overall, it corroborates our model of a thermal boundary layer being confined by the free-surface. 

In conclusion, in this Letter the different dynamic regimes of a freezing rivulet: homogeneous growth and permanent regime, are investigated experimentally and modeled using classical heat transfer theory. After an initial growth where the dynamics is similar to the classical Stefan problem with no influence of the flow, the ice layer thickness saturates to form a very specific structure. This structure is mainly linear and we were able to recover this feature theoretically, the key mechanism being the confinement of a thermal boundary layer in the water rivulet. This original experiment, therefore, constitutes a model system allowing us to easily investigate a broad range of dynamical regimes and to progress in the understanding of the multiphysics aspects of phase changes in capillary flows.\\

We thank D. Mottin for preliminary experiments. A. H. and T. S. thank J. Cobos for fruitful discussions. We thank the Direction G\'en\'erale de l'Armement (DGA) for financial support.

\bibliography{apssamp}

\end{document}